\documentclass[twocolumn]{aastex631} 

\usepackage{amsmath}
\usepackage{booktabs, makecell, longtable}

\newcommand{\aco}{$\alpha_\mathrm{CO}$}
\defcitealias{co-to-h2}{B13}
\defcitealias{2023ApJ...950..119T}{T23}
\defcitealias{2023arXiv231100407C}{C23}

\shortauthors{Teng et al.}

\begin{document}


\title{Star Formation Efficiency in Nearby Galaxies Revealed with a New CO-to-H$_2$ Conversion Factor Prescription}

\newcommand{\UCSD}{\affiliation{Center for Astrophysics and Space Sciences, Department of Physics, University of California San Diego, \\ 9500 Gilman Drive, La Jolla, CA 92093, USA}}

\newcommand{\McMaster}{\affiliation{Department of Physics and Astronomy, McMaster University, 1280 Main Street West, Hamilton, ON L8S 4M1, Canada}}

\newcommand{\CITA}{\affiliation{Canadian Institute for Theoretical Astrophysics (CITA), University of Toronto, 60 St George Street, Toronto, ON M5S 3H8, Canada}}

\newcommand{\Princeton}{\affiliation{Department of Astrophysical Sciences, Princeton University, 4 Ivy Lane, Princeton, NJ 08544, USA}}

\newcommand{\OSU}{\affiliation{Department of Astronomy, The Ohio State University, 140 West 18th Avenue, Columbus, OH 43210, USA}}

\newcommand{\ANU}{\affiliation{Research School of Astronomy and Astrophysics, Australian National University, Canberra, ACT 2611, Australia}}

\newcommand{\ASIAA}{\affiliation{Institute of Astronomy and Astrophysics, Academia Sinica, No. 1, Sec. 4, Roosevelt Road, Taipei 10617, Taiwan}}

\newcommand{\Bonn}{\affiliation{Argelander-Institut f\"ur Astronomie, Universit\"at Bonn, Auf dem H\"ugel 71, 53121 Bonn, Germany}}

\newcommand{\Heidelberg}{\affiliation{Astronomisches Rechen-Institut, Zentrum f\"{u}r Astronomie der Universit\"{a}t Heidelberg, M\"{o}nchhofstra\ss e 12-14, D-69120 Heidelberg, Germany}}

\newcommand{\ITA}{\affiliation{Universit\"{a}t Heidelberg, Zentrum f\"{u}r Astronomie, Institut f\"{u}r Theoretische Astrophysik, \\ Albert-Ueberle-Str 2, D-69120 Heidelberg, Germany}}

\newcommand{\Leiden}{\affiliation{Leiden Observatory, Leiden University, P.O. Box 9513, 2300 RA Leiden, The Netherlands}}

\newcommand{\Maryland}{\affiliation{Department of Astronomy, University of Maryland, College Park, MD 20742, USA}}

\newcommand{\MPE}{\affiliation{Max-Planck-Institut f\"{u}r extraterrestrische Physik, Giessenbachstra{\ss}e 1, D-85748 Garching, Germany}}

\newcommand{\MPIA}{\affiliation{Max-Planck-Institut f\"{u}r Astronomie, K\"{o}nigstuhl 17, D-69117, Heidelberg, Germany}}

\newcommand{\NRAO}{\affiliation{National Radio Astronomy Observatory, 520 Edgemont Road, Charlottesville, VA 22903-2475, USA}}

\newcommand{\OAN}{\affiliation{Observatorio Astron\'{o}mico Nacional (IGN), C/Alfonso XII, 3, E-28014 Madrid, Spain}}

\newcommand{\Tamkang}{\affiliation{Department of Physics, Tamkang University, No.151, Yingzhuan Rd., Tamsui Dist., New Taipei City 251301, Taiwan}}

\newcommand{\Toledo}{\affiliation{Department of Physics and Astronomy, University of Toledo, Ritter Obs., MS \#113, Toledo, OH 43606, USA}}

\newcommand{\Oxford}{\affiliation{Sub-department of Astrophysics, Department of Physics, University of Oxford, Keble Road, Oxford OX1 3RH, UK}}

\newcommand{\UTA}{\affiliation{Instituto de Alta Investigación, Universidad de Tarapacá, Casilla 7D, Arica, Chile}}

\newcommand{\UGent}{\affiliation{Sterrenkundig Observatorium, Universiteit Gent, Krijgslaan 281 S9, B-9000 Gent, Belgium}}

\correspondingauthor{Yu-Hsuan Teng}
\email{yuteng@ucsd.edu}

\author[0000-0003-4209-1599]{Yu-Hsuan Teng}
\UCSD

\author[0000-0003-2551-7148]{I-Da Chiang}
\ASIAA

\author[0000-0002-4378-8534]{Karin M. Sandstrom}
\UCSD
\affiliation{Department of Astronomy \& Astrophysics, University of California San Diego, 9500 Gilman Drive, La Jolla, CA 92093, USA}

\author[0000-0003-0378-4667]{Jiayi Sun}
\McMaster
\CITA
\Princeton

\author[0000-0002-2545-1700]{Adam K. Leroy}
\OSU

\author[0000-0002-5480-5686]{Alberto D. Bolatto}
\Maryland

\author[0000-0003-1242-505X]{Antonio Usero}
\OAN

\author[0000-0002-0509-9113]{Eve C. Ostriker}
\Princeton

\author[0000-0002-0472-1011]{Miguel Querejeta}
\OAN

\author[0000-0002-5235-5589]{J\'er\'emy~Chastenet}
\UGent

\author[0000-0003-0166-9745]{Frank Bigiel}
\Bonn

\author[0000-0003-0946-6176]{M\'ed\'eric~Boquien}
\UTA

\author[0000-0002-8760-6157]{Jakob den Brok}
\affiliation{Center for Astrophysics $\mid$ Harvard \& Smithsonian, 60 Garden St., 02138 Cambridge, MA, USA}

\author[0000-0001-5301-1326]{Yixian Cao}
\MPE

\author[0000-0002-5635-5180]{M\'elanie Chevance}
\ITA
\affiliation{Cosmic Origins Of Life (COOL) Research DAO, coolresearch.io}

\author[0000-0001-8241-7704]{Ryan Chown}
\OSU

\author[0000-0001-6498-2945]{Dario Colombo}
\Bonn

\author[0000-0002-1185-2810]{Cosima Eibensteiner}
\Bonn

\author[0000-0001-6708-1317]{Simon C.~O. Glover}
\ITA

\author[0000-0002-3247-5321]{Kathryn Grasha}
\altaffiliation{ARC DECRA Fellow}
\affiliation{Research School of Astronomy and Astrophysics, Australian National University, Canberra, ACT 2611, Australia} 
\affiliation{ARC Centre of Excellence for All Sky Astrophysics in 3 Dimensions (ASTRO 3D), Australia}   
\affiliation{Visiting Fellow, Harvard-Smithsonian Center for Astrophysics, 60 Garden Street, Cambridge, MA 02138, USA}

\author[0000-0001-9656-7682]{Jonathan~D.~Henshaw}
\affiliation{Astrophysics Research Institute, Liverpool John Moores University, 146 Brownlow Hill, Liverpool L3 5RF, UK}
\MPIA

\author[0000-0002-9165-8080]{Mar\'ia J. Jim\'enez-Donaire}
\OAN
\affiliation{Centro de Desarrollos Tecnológicos, Observatorio de Yebes (IGN), 19141 Yebes, Guadalajara, Spain}

\author[0000-0001-9773-7479]{Daizhong Liu}
\MPE

\author[0000-0001-7089-7325]{Eric J. Murphy}
\NRAO

\author[0000-0002-1370-6964]{Hsi-An Pan}
\affiliation{Department of Physics, Tamkang University, No.151, Yingzhuan Road, Tamsui District, New Taipei City 251301, Taiwan} 

\author[0000-0002-9333-387X]{Sophia K. Stuber}
\MPIA

\author[0000-0002-0012-2142]{Thomas G. Williams}
\Oxford

\begin{abstract}

Determining how galactic environment, especially the high gas densities and complex dynamics in bar-fed galaxy centers, alters the star formation efficiency (SFE) of molecular gas is critical to understanding galaxy evolution. However, these same physical or dynamical effects also alter the emissivity properties of CO, leading to variations in the CO-to-H$_2$ conversion factor (\aco) that impact the assessment of the gas column densities and thus of the SFE. To address such issues, we investigate the dependence of \aco~on local CO velocity dispersion at 150~pc scales using a new set of dust-based \aco~measurements, and propose a new \aco~prescription that accounts for CO emissivity variations across galaxies. Based on this prescription, we estimate the SFE in a sample of 65 galaxies from the PHANGS--ALMA survey. We find increasing SFE towards high surface density regions like galaxy centers, while using a constant or metallicity-based \aco~results in a more homogeneous SFE throughout the centers and disks. Our prescription further reveals a mean molecular gas depletion time of 700~Myr in the centers of barred galaxies, which is overall 3--4 times shorter than in non-barred galaxy centers or the disks. Across the galaxy disks, the depletion time is consistently around 2--3~Gyr regardless of the choice of \aco~prescription. All together, our results suggest that the high level of star formation activity in barred centers is not simply due to an increased amount of molecular gas but also an enhanced SFE compared to non-barred centers or disk regions. 

\end{abstract}


\section{Introduction} \label{sec:intro}

Star formation in galaxies is governed by the amount of molecular gas and the efficiency with which that gas is converted into stars. To understand the evolutionary process of star formation activity within galaxies, it is critical to measure the molecular gas star formation efficiency (SFE; defined as the ratio between star formation rate, SFR, and molecular gas mass, $M_\mathrm{mol}$), or molecular gas depletion time ($t_\mathrm{dep} = 1/\mathrm{SFE}$) \citep[see review by][]{2022ARA&A..60..319S}. Previous studies have found that SFR and molecular gas surface densities are highly correlated \citep[i.e., the molecular Kennicutt-Schmidt relation, or mKS relation;][]{1998ApJ...498..541K} and that $t_\mathrm{dep}$ is usually at 1--4~Gyr across nearby star-forming galaxies \citep[e.g.,][]{2008AJ....136.2846B,2008AJ....136.2782L,2011MNRAS.415...61S,stacking,2017ApJ...849...26U,2023ApJ...945L..19S}. Despite the minor variation in general, $t_\mathrm{dep}$ is also found to vary systematically with local and global host galaxy properties, which could be driven by environmental and/or dynamical effects from e.g., metallicity, molecular cloud structure, bar instabilities, active galactic nuclei, or galaxy interactions \citep{2011MNRAS.415...61S,2012ApJ...758...73S,2019ApJ...883....2S,2021MNRAS.501.4777E,2021MNRAS.505L..46E,2021A&A...656A.133Q,2022MNRAS.514.5035L,2022ApJ...940..176V,2023A&A...671A...3J,2023ApJ...943....7M}.

The assessment of molecular gas SFE relies heavily on the CO-to-H$_2$ conversion factor (\aco)\footnote{\aco~is defined for the CO $J=1-0$ line in most literature, but it can also be evaluated for other transitions. In this work, when we refer to \aco, we mean $\alpha_\mathrm{CO(1-0)}$ unless otherwise specified.}: 
\begin{equation}
\alpha_\mathrm{CO} = \frac{M_\mathrm{mol}}{L'_\mathrm{CO(1-0)}} = \frac{\Sigma_\mathrm{mol}}{I_\mathrm{CO(1-0)}}\ \rm \left[\frac{M_\odot}{K\ km~s^{-1}\ pc^2} \right]~,
\label{def_alphaCO}
\end{equation}
where $M_\mathrm{mol}$ ($\Sigma_\mathrm{mol}$) is the total molecular gas mass (surface density) and $L'_\mathrm{CO(1-0)}$ ($I_\mathrm{CO(1-0)}$) is the line luminosity (intensity) of CO $J$=1--0. \aco~is known to vary with molecular gas conditions such as density, temperature, and dynamical state \citep[see review by][hereafter \citetalias{co-to-h2}]{co-to-h2}, which are the same conditions that could also alter the intrinsic SFE of the molecular gas. Due to the lack of a widely-agreed prescription that can accurately predict \aco, many studies could only assume a constant \aco~referencing the Milky Way (MW) disk average (e.g., \citetalias{co-to-h2}) to convert CO observations to molecular gas mass. This has made \aco~variation one of the dominant sources of uncertainty in current molecular gas and SFE studies \citep[see discussions in][]{2020MNRAS.492.6027E,2023ApJ...943....7M,2023ApJ...945L..19S}.

The impacts of \aco~variations on both SFE and cloud evolutionary timescale estimates are particularly critical in galaxy centers \citep{2013AJ....146...19L,2017ApJ...849...26U,2019PASJ...71S..15M,2020MNRAS.492.6027E,2021A&A...650A.134P,2023ApJ...943....7M,2023ApJ...945L..19S}.
In those environments, \aco~can be 5--15 times lower than the Galactic disk value \citep{2012ApJ...750....3A,2013ApJ...777....5S,2020A&A...635A.131I,2022ApJ...925...72T,2023ApJ...950..119T,2023arXiv230203044D}. The lower \aco~in galaxy centers is likely driven by CO emissivity variations due to higher excitation and/or stronger dynamical effects such as turbulence or inflowing gas \citep{2012MNRAS.421.3127N,2012ApJ...751...10P,co-to-h2,2020ApJ...903..142G,2023ApJ...950..119T}. These effects may also explain the low \aco~seen in mergers or (ultra-)luminous infrared galaxies (U/LIRGs) \citep{1998ApJ...507..615D,2017ApJ...850...77K,2017ApJ...840....8S,2018ApJ...863..143C,2019A&A...628A..71H}.  

Reducing the uncertainty in molecular gas and SFE studies, and thereby improving our understanding in star formation and galaxy evolution, requires a robust \aco~prescription that can be systematically applied to large samples of galaxies with diverse environments. Recent studies have proposed various types of \aco~prescription depending on metallicity, stellar mass surface density, SFR, SFE, and/or CO line-related properties \citep{2012ApJ...746...69G,2012MNRAS.421.3127N,co-to-h2,2015A&A...583A.114H,2016A&A...588A..23A,2017MNRAS.464.3315A,2019A&A...621A.104R,2020ApJ...903..142G,2020A&A...643A.141M,2023arXiv230614881R}. 
However, establishing a reliable \aco~calibration remains a challenge because it requires \aco~measurements covering a sufficient sample of galaxies spanning a broad range of molecular gas physical and dynamical conditions, and the two most realistic ways to measure \aco~in nearby galaxies are via dust emission (which is typically restricted to kpc resolutions; \citealt{1997A&A...328..471I,2011ApJ...737...12L,2013ApJ...777....5S,2017ApJ...835..278S,2021ApJS..256....3P,2023arXiv230203044D,2023PASJ...75..743Y,2023arXiv231100407C}) or multi-CO isotopologue observations (which is expensive at cloud scales; \citealt{2017ApJ...840....8S,2018MNRAS.475.3909C,2020A&A...635A.131I,2022MNRAS.509.2180S,2022ApJ...925...72T,2023ApJ...950..119T}).

Thanks to the high resolution and sensitivity of the Atacama Large Millimeter/submillimeter Array (ALMA), CO (isotopologue) observations are now routinely possible at cloud scales in nearby galaxies \citep[e.g.,][]{2021ApJS..257...43L,2022MNRAS.512.1522D,2023ApJ...949..108K,2023MNRAS.525.4270W}. In particular, recent studies modeling multi-CO isotopologues in nearby galaxy centers have revealed that CO opacity is the dominant driver of \aco~variations \citep{2020A&A...635A.131I,2022ApJ...925...72T,2023ApJ...950..119T}. This strong dependence of \aco~on CO opacity further leads to a clear anti-correlation between \aco~and the observed line width at $\sim$100~pc scales in barred galaxy centers \citep[][hereafter \citetalias{2023ApJ...950..119T}]{2023ApJ...950..119T}.

Motivated by these latest measurements of \aco, we will test if the correlation found in \citetalias{2023ApJ...950..119T} also applies to the 12 galaxies (labeled with * in Table~\ref{tab:sample}, including 8 barred and 4 non-barred) which have dust-inferred \aco~values at kpc scales (from \citealt{2023arXiv231100407C}; hereafter \citetalias{2023arXiv231100407C}) and molecular gas velocity dispersion measured at 150-pc scales \citep[from the PHANGS--ALMA survey;][]{2021ApJS..257...43L,2022AJ....164...43S}. The results of this comparison lead us to a new \aco~prescription capturing CO emissivity effects in star-forming galaxies. In this paper, we present this prescription, discuss its physical implications, and study its impact on SFE across a sizable sample of galaxy centers and disks with diverse properties.

\section{Data and Measurements} \label{sec:data}

\subsection{PHANGS Datasets}

Our analysis is based on various molecular gas and star formation properties, leveraging a database developed by \citet{2022AJ....164...43S} which assembled multi-wavelength measurements of 80 galaxies from the PHANGS--ALMA survey \citep{2021ApJS..257...43L}. From this database, we extract multiple physical quantities in matched hexagonal apertures with fixed sizes of 1.5~kpc. The quantities used in this work include: intensity-weighted mean molecular gas velocity dispersion measured at 150-pc scale ($\langle \Delta v \rangle_\mathrm{150pc}$), area-weighted mean CO(2--1) line integrated intensity ($I_\mathrm{CO(2-1)}$), stellar mass surface density ($\Sigma_\mathrm{star}$), SFR surface density ($\Sigma_\mathrm{SFR}$), and gas-phase metallicity ($Z'$, normalized to the solar value $[12+\log(\mathrm{O/H})_\odot = 8.69]$ and calibrated based on \citealt{2004MNRAS.348L..59P}). All these quantities are corrected for the effects of galaxy inclination and data sensitivity limits (see \citealt{2022AJ....164...43S} for more details).

To further explore trends in galaxies with or without stellar bars, we adopt the classification of stellar bars for PHANGS galaxies \citep{2021A&A...656A.133Q}. Table~\ref{tab:sample} lists the 65 galaxies included in our analysis, which is the overlap between \citet{2021A&A...656A.133Q} and \citet{2022AJ....164...43S}. This sample from PHANGS has high-resolution CO(2--1) data with beam sizes of 150~pc or smaller. Columns (6--10) in Table~\ref{tab:sample} show the measurements extracted from \citet{2022AJ....164...43S} for the central 1.5 kpc regions of those galaxies.

\begin{longtable*}{lccccccccccc}
\caption{Galaxy Sample and Properties in the Central 1.5 kpc Regions \label{tab:sample}} \\
\hline \hline
   Galaxy & Bar & Dist. & Incl. &   P.A. & $Z'$ & $\log(\Sigma_\mathrm{SFR})$ & $I_\mathrm{CO(2-1)}$ & $\log(\Sigma_\mathrm{star})$ & $\langle \Delta v \rangle_\mathrm{150pc}$ & $\log(\alpha_\mathrm{CO}^\mathrm{Eq.\ref{eqn_ew21_fit}})$ & $\log(t_\mathrm{dep})$ \\
   &  & [Mpc] & [deg] & [deg] & [Z$_\odot$] & $\left[ \frac{\mathrm{M_\odot}}{\mathrm{yr~kpc^2}} \right]$ & [K~km~s$^{-1}$] & [M$_\odot$ pc$^{-2}$] & [km~s$^{-1}$] & $\left[ \frac{\mathrm{M_\odot\ s}}{\mathrm{K~km~pc^2}} \right]$ & [yr] \\
   (1) & (2) & (3) & (4) & (5) & (6) & (7) & (8) & (9) & (10) & (11) & (12) \\ 
\hline
\endfirsthead

\toprule
   Galaxy & Bar & Dist. & Incl. &   P.A. & $Z'$ & $\log(\Sigma_\mathrm{SFR})$ & $I_\mathrm{CO(2-1)}$ & $\log(\Sigma_\mathrm{star})$ & $\langle \Delta v \rangle_\mathrm{150pc}$ & $\log(\alpha_\mathrm{CO}^\mathrm{Eq.\ref{eqn_ew21_fit}})$ & $\log(t_\mathrm{dep})$ \\
   &  & [Mpc] & [deg] & [deg] & [Z$_\odot$] & $\left[ \frac{\mathrm{M_\odot}}{\mathrm{yr~kpc^2}} \right]$ & [K~km~s$^{-1}$] & [M$_\odot$ pc$^{-2}$] & [km~s$^{-1}$] & $\left[ \frac{\mathrm{M_\odot\ s}}{\mathrm{K~km~pc^2}} \right]$ & [yr] \\
   (1) & (2) & (3) & (4) & (5) & (6) & (7) & (8) & (9) & (10) & (11) & (12) \\
\midrule
\endhead
\midrule
\multicolumn{12}{r}{{Continued on next page}} \\
\endfoot

\bottomrule \\
\multicolumn{12}{l}{\parbox{\dimexpr\textwidth-2\tabcolsep}{%
     \textbf{Note.} (1) Galaxies with an asterisk are those with \aco~measurements (see Section~\ref{subsec:data_aco}); (2) bar classification \citep{2021A&A...656A.133Q}; (3) distance \citep{2021MNRAS.501.3621A}; (4--5) inclination and position angles \citep{2020ApJ...897..122L}; (6--10) the central 1.5~kpc measurements of gas-phase metallicity (PP04-based), kpc-averaged SFR surface density, kpc-averaged CO(2-1) integrated intensity, kpc-averaged stellar mass surface density, and CO intensity-weighted mean velocity dispersion at 150-pc scale \citep{2022AJ....164...43S}; (11) $\log(\alpha_\mathrm{CO})$ derived from (10) using Equation~\ref{eqn_ew21_fit}; (12) molecular gas depletion time derived from (7), (8), and (11) using Equation~\ref{eqn_sfe}.}} \\
\endlastfoot
  IC1954 &   1 &  12.8 &  57.1 &  63.4 & 1.10 &                       -1.67 &                  7.5 &                         2.51 &                                       7.1 &                                     0.36 &                   9.08 \\
  IC5273 &   1 &  14.2 &  52.0 & 234.1 & 1.12 &                       -1.59 &                  4.7 &                         2.48 &                                       7.4 &                                     0.34 &                   8.78 \\
 NGC0253* &   1 &   3.7 &   75.0 &  52.5 & 1.33 &                0.3166 &                  198.2 &                 990.2 &                                       28.5 &                                     -0.15 &                   8.88 \\
 NGC0628* &   0 &   9.8 &   8.9 &  20.7 & 1.29 &                       -1.82 &                  6.3 &                         3.03 &                                       5.4 &                                     0.45 &                   9.26 \\
 NGC0685 &   1 &  19.9 &  23.0 & 100.9 & 1.23 &                       -2.15 &                  2.8 &                         2.42 &                                       6.8 &                                     0.37 &                   9.15 \\
 NGC1087 &   1 &  15.8 &  42.9 & 359.1 & 1.19 &                       -1.12 &                 24.7 &                         2.68 &                                      13.4 &                                     0.12 &                   8.82 \\
 NGC1097 &   1 &  13.6 &  48.6 & 122.4 & 1.34 &                       -0.29 &                196.0 &                         3.61 &                                      33.1 &                                    -0.20 &                   8.57 \\
 NGC1300 &   1 &  19.0 &  31.8 & 278.0 & 1.33 &                       -1.57 &                 44.6 &                         3.25 &                                      23.2 &                                    -0.07 &                   9.33 \\
 NGC1317 &   1 &  19.1 &  23.2 & 221.5 & 1.33 &                       -1.45 &                 25.2 &                         3.64 &                                      18.0 &                                     0.02 &                   9.06 \\
 NGC1365 &   1 &  19.6 &  55.4 & 201.1 & 1.36 &                        0.04 &                462.4 &                         3.80 &                                      25.6 &                                    -0.11 &                   8.70 \\
 NGC1385 &   0 &  17.2 &  44.0 & 181.3 & 1.21 &                       -1.00 &                 15.0 &                         2.78 &                                       9.5 &                                     0.25 &                   8.60 \\
 NGC1433 &   1 &  18.6 &  28.6 & 199.7 & 1.35 &                       -1.46 &                 44.6 &                         3.63 &                                      17.3 &                                     0.03 &                   9.33 \\
 NGC1511 &   0 &  15.3 &  72.7 & 297.0 & 1.18 &                       -1.17 &                 12.4 &                         2.60 &                                       7.5 &                                     0.33 &                   8.79 \\
 NGC1512 &   1 &  18.8 &  42.5 & 261.9 & 1.34 &                       -1.36 &                 31.9 &                         3.41 &                                      11.3 &                                     0.19 &                   9.24 \\
 NGC1546 &   0 &  17.7 &  70.3 & 147.8 & 1.29 &                       -1.52 &                 45.0 &                         3.08 &                                       7.9 &                                     0.31 &                   9.67 \\
 NGC1559 &   1 &  19.4 &  65.4 & 244.5 & 1.29 &                       -1.47 &                 10.5 &                         2.77 &                                       6.6 &                                     0.38 &                   9.06 \\
 NGC1566 &   1 &  17.7 &  29.5 & 214.7 & 1.34 &                       -0.97 &                 55.7 &                         3.63 &                                      28.7 &                                    -0.15 &                   8.76 \\
 NGC1637 &   1 &  11.7 &  31.1 &  20.6 & 1.20 &                       -0.82 &                 12.2 &                         2.57 &                                      22.8 &                                    -0.07 &                   8.03 \\
 NGC1792 &   0 &  16.2 &  65.1 & 318.9 & 1.33 &                       -1.35 &                 52.0 &                         3.08 &                                      10.7 &                                     0.21 &                   9.46 \\
 NGC1809 &   0 &  20.0 &  57.6 & 138.2 & 1.14 &                       -1.71 &                  2.2 &                         2.19 &                                       3.1 &                                     0.65 &                   8.88 \\
 NGC2090 &   0 &  11.8 &  64.5 & 192.5 & 1.22 &                       -2.19 &                  4.5 &                         2.79 &                                       4.2 &                                     0.54 &                   9.57 \\
 NGC2283 &   1 &  13.7 &  43.7 &  -4.1 & 1.18 &                       -1.93 &                  3.0 &                         2.31 &                                       5.1 &                                     0.47 &                   9.07 \\
 NGC2566 &   1 &  23.4 &  48.5 & 312.0 & 1.34 &                       -0.05 &                265.0 &                         3.16 &                                      26.5 &                                    -0.12 &                   8.53 \\
 NGC2835 &   1 &  12.2 &  41.3 &   1.0 & 1.21 &                       -2.23 &                  2.2 &                         2.47 &                                       4.3 &                                     0.53 &                   9.30 \\
 NGC2903 &   1 &  10.0 &  66.8 & 203.7 & 1.33 &                       -0.81 &                 55.8 &                         3.15 &                                      19.7 &                                    -0.01 &                   8.73 \\
 NGC2997 &   0 &  14.1 &  33.0 & 108.1 & 1.34 &                       -0.95 &                 68.5 &                         3.21 &                                      16.2 &                                     0.06 &                   9.03 \\
 NGC3059 &   1 &  20.2 &  29.4 & -14.8 & 1.29 &                       -0.84 &                 23.1 &                         2.68 &                                      11.4 &                                     0.18 &                   8.58 \\
 NGC3137 &   0 &  16.4 &  70.3 &  -0.3 & 1.18 &                       -2.53 &                  3.0 &                         2.19 &                                       3.2 &                                     0.64 &                   9.85 \\
 NGC3351* &   1 &  10.0 &  45.1 & 193.2 & 1.29 &                       -0.71 &                 34.4 &                         3.18 &                                      19.2 &                                    -0.01 &                   8.43 \\
 NGC3507 &   1 &  23.6 &  21.7 &  55.8 & 1.30 &                       -1.57 &                 17.3 &                         3.16 &                                      23.4 &                                    -0.08 &                   8.91 \\
 NGC3511 &   1 &  13.9 &  75.1 & 256.8 & 1.22 &                       -1.94 &                 17.2 &                         2.50 &                                       6.6 &                                     0.38 &                   9.74 \\
 NGC3521* &   0 &  13.2 &  68.8 & 343.0 & 1.36 &                       -1.66 &                 14.8 &                         3.50 &                                       6.7 &                                     0.37 &                   9.40 \\
 NGC3596 &   0 &  11.3 &  25.1 &  78.4 & 1.10 &                       -1.55 &                  6.3 &                         2.68 &                                       7.0 &                                     0.36 &                   8.90 \\
 NGC3621 &   0 &   7.1 &  65.8 & 343.8 & 1.23 &                       -1.86 &                  8.4 &                         2.66 &                                       5.2 &                                     0.46 &                   9.44 \\
 NGC3626 &   1 &  20.0 &  46.6 & 165.2 & 1.31 &                       -1.40 &                 11.2 &                         3.52 &                                      13.5 &                                     0.12 &                   8.76 \\
 NGC3627* &   1 &  11.3 &  57.3 & 173.1 & 1.35 &                       -1.19 &                 64.9 &                         3.53 &                                      34.6 &                                    -0.22 &                   8.98 \\
 NGC4254* &   0 &  13.1 &  34.4 &  68.1 & 1.30 &                       -1.10 &                 35.7 &                         3.27 &                                      10.3 &                                     0.22 &                   9.06 \\
 NGC4293 &   1 &  15.8 &  65.0 &  48.3 & 1.31 &                       -1.11 &                 45.0 &                         2.92 &                                      27.4 &                                    -0.13 &                   8.82 \\
 NGC4298 &   0 &  14.9 &  59.2 & 313.9 & 1.22 &                       -1.79 &                 14.3 &                         2.69 &                                       9.7 &                                     0.24 &                   9.38 \\
 NGC4303 &   1 &  17.0 &  23.5 & 312.4 & 1.32 &                       -0.83 &                 70.8 &                         3.52 &                                      16.9 &                                     0.04 &                   8.91 \\
 NGC4321* &   1 &  15.2 &  38.5 & 156.2 & 1.34 &                       -0.78 &                101.0 &                         3.35 &                                      19.5 &                                    -0.01 &                   8.96 \\
 NGC4457 &   1 &  15.1 &  17.4 &  78.7 & 1.30 &                       -1.15 &                 38.7 &                         3.69 &                                      29.0 &                                    -0.15 &                   8.77 \\
NGC4496A &   1 &  14.9 &  53.8 &  51.1 & 1.04 &                       -2.21 &                  1.6 &                         2.17 &                                       2.8 &                                     0.69 &                   9.29 \\
 NGC4535 &   1 &  15.8 &  44.7 & 179.7 & 1.32 &                       -1.03 &                 43.1 &                         2.78 &                                      20.1 &                                    -0.02 &                   8.83 \\
 NGC4536* &   1 &  16.2 &  66.0 & 305.6 & 1.30 &                       -0.56 &                110.3 &                         3.30 &                                      21.0 &                                    -0.04 &                   8.76 \\
 NGC4540 &   1 &  15.8 &  28.7 &  12.8 & 1.14 &                       -1.90 &                  3.9 &                         2.65 &                                       6.3 &                                     0.40 &                   9.07 \\
 NGC4548 &   1 &  16.2 &  38.3 & 138.0 & 1.34 &                       -1.96 &                  9.7 &                         3.38 &                                      24.8 &                                    -0.10 &                   9.03 \\
 NGC4569* &   1 &  15.8 &  70.0 &  18.0 & 1.35 &                       -1.13 &                112.1 &                         3.24 &                                      27.6 &                                    -0.14 &                   9.23 \\
 NGC4571 &   0 &  14.9 &  32.7 & 217.5 & 1.23 &                       -2.36 &                  2.5 &                         2.63 &                                       2.7 &                                     0.70 &                   9.63 \\
 NGC4689* &   0 &  15.0 &  38.7 & 164.1 & 1.26 &                       -1.81 &                  9.2 &                         2.61 &                                       5.6 &                                     0.44 &                   9.40 \\
 NGC4731 &   1 &  13.3 &  64.0 & 255.4 & 1.02 &                       -1.97 &                  1.6 &                         1.99 &                                       5.1 &                                     0.47 &                   8.85 \\
 NGC4781 &   1 &  11.3 &  59.0 & 290.0 & 1.09 &                       -1.76 &                  9.1 &                         2.59 &                                       8.2 &                                     0.30 &                   9.20 \\
 NGC4826 &   0 &   4.4 &  59.1 & 293.6 & 1.27 &                       -1.43 &                 27.9 &                         3.26 &                                      21.8 &                                    -0.05 &                   9.01 \\
 NGC4941* &   1 &  15.0 &  53.4 & 202.2 & 1.25 &                       -1.31 &                  6.9 &                         3.00 &                                      18.7 &                                     0.00 &                   8.34 \\
 NGC4951 &   0 &  15.0 &  70.2 &  91.2 & 1.15 &                       -1.77 &                 10.0 &                         2.67 &                                      11.4 &                                     0.18 &                   9.14 \\
 NGC5042 &   0 &  16.8 &  49.4 & 190.6 & 1.18 &                       -2.15 &                  2.2 &                         2.52 &                                       5.3 &                                     0.46 &                   9.14 \\
 NGC5068 &   1 &   5.2 &  35.7 & 342.4 & 0.98 &                       -2.01 &                  1.0 &                         2.33 &                                       3.6 &                                     0.60 &                   8.79 \\
 NGC5128 &   0 &   3.7 &  45.3 &  32.2 & 1.36 &                       -0.78 &                 45.3 &                         3.70 &                                      22.7 &                                    -0.07 &                   8.56 \\
 NGC5134 &   1 &  19.9 &  22.7 & 311.6 & 1.30 &                       -1.94 &                  4.9 &                         3.33 &                                      11.4 &                                     0.18 &                   9.00 \\
 NGC5248* &   1 &  14.9 &  47.4 & 109.2 & 1.30 &                       -0.97 &                 72.6 &                         3.35 &                                      15.7 &                                     0.07 &                   9.08 \\
 NGC5530 &   0 &  12.3 &  61.9 & 305.4 & 1.23 &                       -2.02 &                  6.4 &                         3.00 &                                       4.8 &                                     0.50 &                   9.51 \\
 NGC5643 &   1 &  12.7 &  29.9 & 318.7 & 1.29 &                       -0.29 &                 42.0 &                         2.88 &                                      26.7 &                                    -0.12 &                   7.98 \\
 NGC6300 &   1 &  11.6 &  49.6 & 105.4 & 1.31 &                       -0.55 &                 41.1 &                         2.85 &                                      36.2 &                                    -0.23 &                   8.11 \\
 NGC7456 &   0 &  15.7 &  67.3 &  16.0 & 1.09 &                       -2.70 &                  0.9 &                         1.99 &                                       2.7 &                                     0.70 &                   9.56 \\
 NGC7496 &   1 &  18.7 &  35.9 & 193.7 & 1.21 &                       -0.43 &                 46.2 &                         2.72 &                                      23.7 &                                    -0.08 &                   8.20 \\
\end{longtable*}

\subsection{Dust-based \aco~Measurements} \label{subsec:data_aco}

We obtain spatially resolved \aco~from \citetalias{2023arXiv231100407C}, where \aco~is measured at 2~kpc resolution across 41 nearby ($\leq 20$~Mpc) and moderately-inclined ($\text{Incl.}\leq 80^\circ$) spiral galaxies with resolved measurements of CO integrated intensity (including PHANGS-ALMA) and atomic gas.
The authors assumed a constant dust-to-metals ratio to constrain the total gas mass with dust and metallicity measurements. In their sample, 8 barred and 4 non-barred galaxies from PHANGS have dust-based \aco\ measurements (those with an * in Table~\ref{tab:sample}). These measurements typically cover out to a galactocentric radius of $\sim$10~kpc, including $\sim$2000 Nyquist-sampled data points. It is based on these data that we examine scaling relations of \aco~and develop an \aco~prescription in Section~\ref{subsec:prescription}.   

The \aco~measurements in \citetalias{2023arXiv231100407C} were derived based on the PHANGS CO(2--1) data, and we directly use their $\alpha_\mathrm{CO(2-1)}$ measurements to ensure methodological consistency when we derive molecular gas surface density and SFE (see Sections~\ref{subsec:sfe_centers} and~\ref{subsec:sfe}). To compare with most \aco~literature using $\alpha_\mathrm{CO(1-0)}$, however, we convert the measured $\alpha_\mathrm{CO(2-1)}$ to $\alpha_\mathrm{CO(1-0)}$ by assuming a CO(2--1)/(1--0) ratio ($R_{21}$) of 0.65. Such results can be easily reverted to $\alpha_\mathrm{CO(2-1)}$ via a linear scaling with 0.65. We note that \citetalias{2023arXiv231100407C} also provided $\alpha_\mathrm{CO(1-0)}$ measurements assuming a SFR-dependent $R_{21}$, and we have checked that using such \aco\ does not change any of our results qualitatively (see Section~\ref{subsec:prescription}). 

We also note that the metallicity adopted by \citetalias{2023arXiv231100407C} for computing \aco~is based on the S-calibration in \citet[hereafter PG16S]{2016MNRAS.457.3678P}, which is different from the O3N2 calibration used for the PHANGS dataset based on \citet[hereafter PP04]{2004MNRAS.348L..59P}. Recent studies suggest that PG16S is a more reliable metallicity prescription than PP04 \citep[e.g.,][]{2019ApJ...887...80K}. With the data on 12 galaxies, we find that PP04 estimates result in $\sim$0.2~dex higher $Z'$ than PG16S \citep[see also][]{2019A&A...623A...5D}, which might be due to the mismatch in the adopted solar oxygen abundance value under different calibration schemes \citep[e.g., $12+\log(\mathrm{O/H})_\odot = 8.50$ or 8.69; see discussion in][]{2022ApJ...931...92E}. Throughout this work, we adopt PG16S-based $Z'$ from \citetalias{2023arXiv231100407C} for analyses restricted to these 12 galaxies. However, due to the lack of PG16S-based measurements on all 65 PHANGS galaxies, we use the PP04-based $Z'$ when implementing metallicity-dependent \aco~prescriptions across the full sample for consistency.       

To evaluate the credibility of the observed \aco~trends with our parameters of interest (i.e., $\langle \Delta v \rangle_\mathrm{150pc}$ and $Z'$), we calculate for each parameter bin the number of pixels with reliable \aco~measurements divided by the number of pixels with measured $\Delta v$ or $Z'$. For $\Delta v$, we find the fraction of reliable pixels to be 70--100\% for bins with $\langle \Delta v \rangle_\mathrm{150pc} \gtrsim 3$ km~s$^{-1}$, while it drops significantly to $< 50\%$ in lower velocity dispersion bins\footnote{This is likely due to a large amount of low S/N measurements clustering around $\sim$2.5 km/s, which is the velocity resolution of the PHANGS CO data}. This means that our \aco~data coverage is insufficient to accurately represent regions with $\Delta v \lesssim 3$ km~s$^{-1}$. As for $Z'$, the corresponding completeness of \aco~is above $60\%$ across regions with $Z' \gtrsim 0.6$, while it drops below $40\%$ at lower metallicities (where the PHANGS-ALMA dataset has poorer coverage).  These ``incomplete'' regimes will be excluded by our fitting and analysis in Section~\ref{subsec:prescription}, where we present the new \aco~prescription.

\section{Results} \label{sec:result}

\begin{figure*}
\begin{minipage}{.49\linewidth}
\centering
\includegraphics[width=\linewidth]{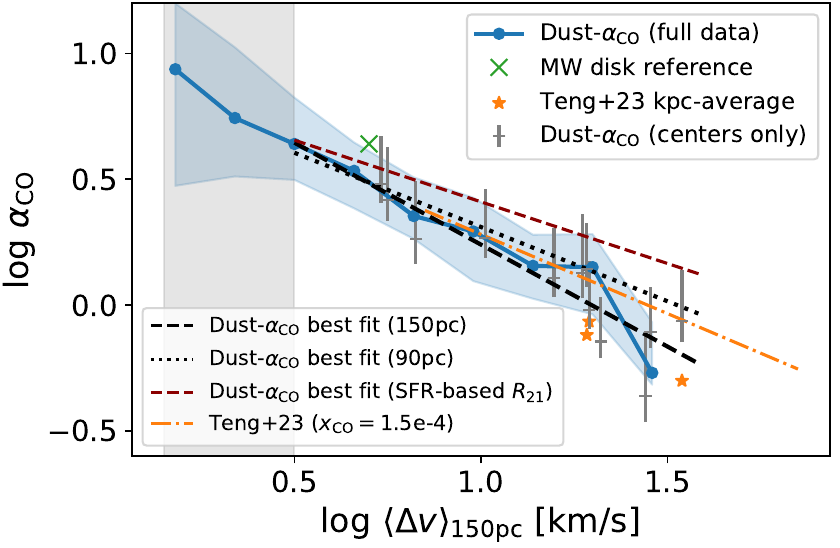}
\end{minipage}
\hfill
\begin{minipage}{.495\linewidth}
\centering
\includegraphics[width=\linewidth]{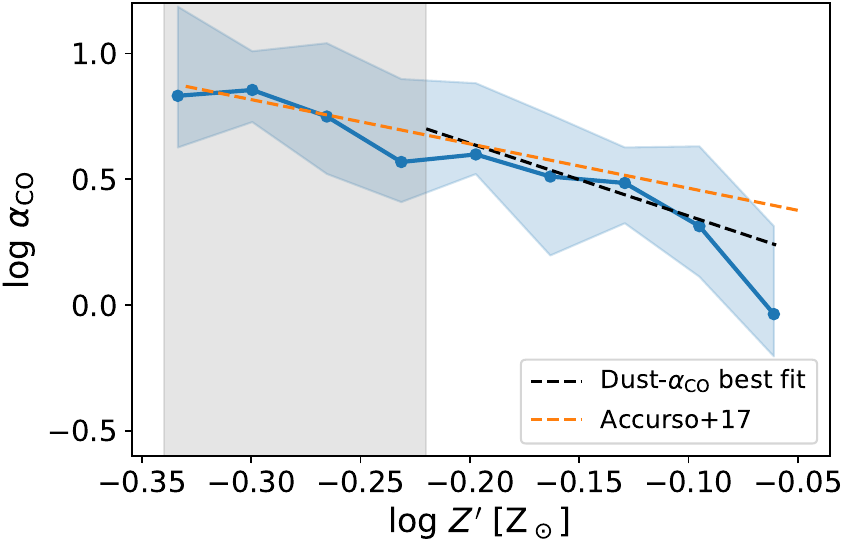}
\end{minipage} \\ \\
\begin{minipage}{.5\linewidth}
\centering
\includegraphics[width=\linewidth]{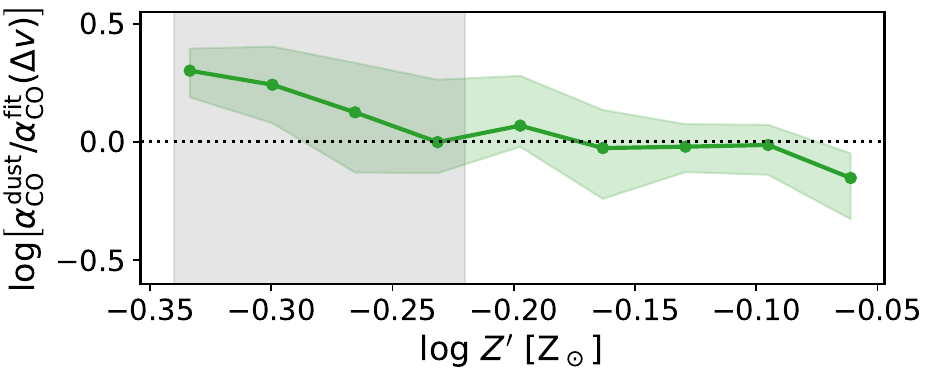}\\
(a)
\end{minipage}
\hfill
\begin{minipage}{.485\linewidth}
\centering
\includegraphics[width=\linewidth]{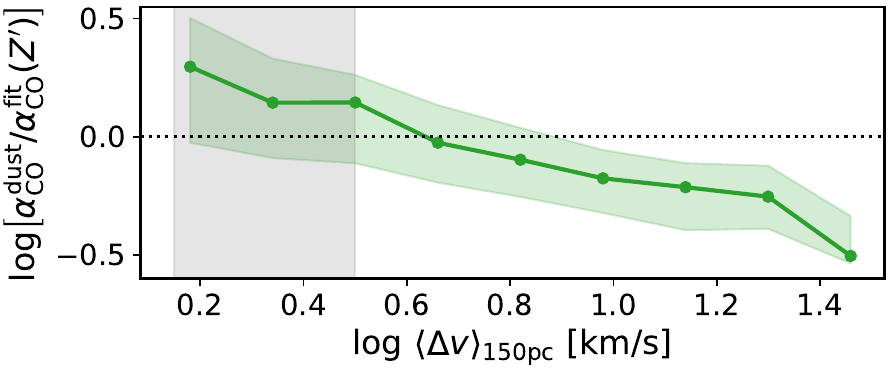}\\
(b)
\end{minipage}
\caption{Column (a): Dust-based \aco~measurements show a strong anti-correlation with the intensity-weighted average of 150-pc scale molecular gas velocity dispersion (top), consistent with the result from \citetalias{2023ApJ...950..119T} on barred galaxy centers (orange line, with an assumed CO/H$_2$ abundance $x_\mathrm{CO} = 1.5\times10^{-4}$); the blue lines and shaded area represent the binned medians and 16th--84th percentile of the measured \aco; the gray shaded area indicates low-confidence regime where \aco~sampling is incomplete; the black dashed/dotted lines show the best fit power-law relations with $\langle \Delta v \rangle$ at 150/90~pc resolutions, and the red dashed line represents the best fit relation when \aco~is derived by assuming a SFR-dependent $R_{21}$ (\citetalias{2023arXiv231100407C}). The residuals of the fit (bottom) do not correlate with $Z'$ in the data-complete regime, suggesting that the observed \aco~variations can be fully captured by our $\Delta v$-based prescription, without requiring an additional metallicity dependence. Column (b): Similar to (a), but the measured \aco~is correlated with metallicity (top), and the residuals are correlated with $\langle \Delta v \rangle_\mathrm{150pc}$ (bottom); the orange dashed line marks the prediction from \citet{2017MNRAS.464.3315A}, which agrees with the overall data but shows a larger scatter.}
\label{fig:dust-aco}
\end{figure*}

\begin{figure*}
\begin{minipage}{.52\linewidth}
\centering
\includegraphics[width=\linewidth]{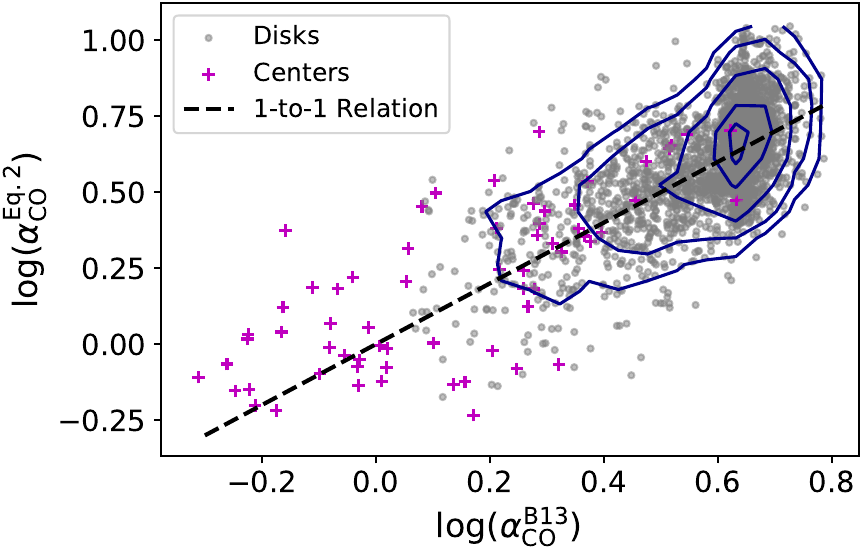}\\
(a)
\end{minipage}
\hfill
\begin{minipage}{.475\linewidth}
\centering
\includegraphics[width=\linewidth]{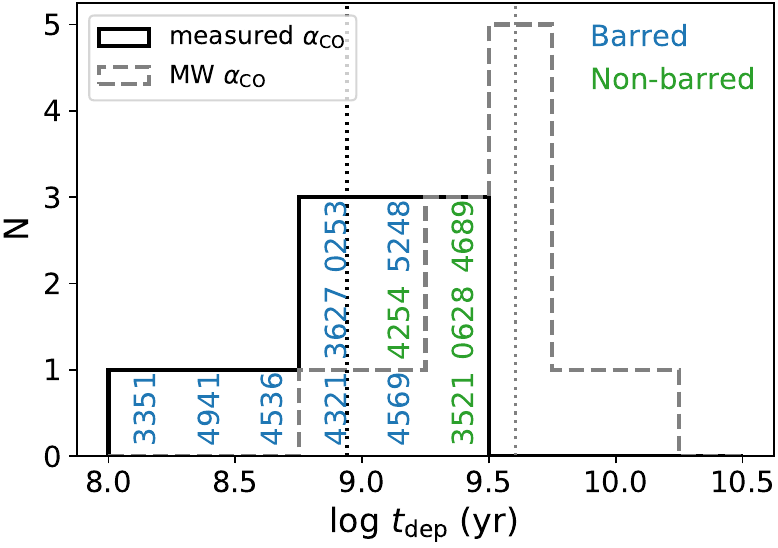}\\
(b)
\end{minipage}
\caption{(a) Comparison of the derived \aco~using our $\Delta v$-based prescription (Equation~\ref{eqn_ew21_fit}) and the $Z'$ plus $\Sigma_\mathrm{star}$-based prescription (Equation~\ref{eqn_alpha_B13}; \citetalias{co-to-h2}), applied to 65 galaxies. The overlaid contours indicate 16\%, 50\%, 84\%, 95\%, and 98\% data inclusion of the disk regions. The two prescriptions show a general 1-to-1 agreement (dashed line), which supports the credibility of our prescription. (b) Molecular gas depletion time ($t_\mathrm{dep}$) of 12 galaxy centers (with their NGC names shown on the histogram) determined by the measured \aco~(solid line) and the MW \aco~(dashed line). The median $t_\mathrm{dep}$ using measured or MW \aco~are indicated by the vertical dotted lines. Overall, $t_\mathrm{dep}$ is lower using the measured \aco, and a clear separation is found between barred and non-barred galaxies, suggesting high star formation efficiency in barred galaxy centers.}
\label{fig:compare-B13}
\end{figure*}

\subsection{A Velocity Dispersion-based \aco~Prescription} \label{subsec:prescription}

To investigate how the dust-based \aco~varies with local velocity dispersion, we use nearest-neighbor matching to relate the \aco~measurements at 2-kpc scales with the velocity dispersion which is measured at 150-pc scale and then averaged over 1.5-kpc-sized apertures via intensity weighting ($\langle \Delta v \rangle_\mathrm{150pc}$). 
As shown in Figure~\ref{fig:dust-aco}(a), the data clearly follow an inverse power-law relation, which is in close agreement with the fit by \citetalias{2023ApJ...950..119T} on three barred galaxy centers at $\sim$100~pc scales (i.e., dash-dotted orange line, assuming a CO/H$_2$ abundance of $1.5\times10^{-4}$). 
The central regions of the 12 galaxies (vertical, gray bars in Figure~\ref{fig:dust-aco}(a)) align well with the overall trend, showing that velocity dispersion can trace \aco~variations in both the centers and disks\footnote{In this paper, \textit{center} refers to the central $\sim$2~kpc-sized aperture at $R_\mathrm{gal}$=0, and \textit{disks} represents the rest of the measurements.}. The green cross sign marks the typical MW disk values of \aco $\sim$4.35 $\mathrm{M_\odot\ (K~km~s^{-1}~pc^2)^{-1}}$ and $\Delta v = 5$~km~s$^{-1}$, which also agrees with the overall trend. The shaded area in Figure~\ref{fig:dust-aco}(a) indicates the regime where \aco~data is incomplete (see Section~\ref{subsec:data_aco}).   

Excluding the incomplete regime, we conduct a least-squares fitting in log-log space based on the remaining $\sim$1600 data points, using the \texttt{curve\_fit} function in \texttt{scipy.optimize}.
The best-fit power-law relation to these data from 12 galaxies is represented by
\begin{equation}
\log \alpha_\mathrm{CO} = -0.81\ \log \langle \Delta v \rangle_\mathrm{150pc} + 1.05~,
\label{eqn_ew21_fit}
\end{equation}
where \aco~and $\langle \Delta v \rangle_\mathrm{150pc}$ are in units of $\mathrm{M_\odot\ (K~km~s^{-1}~pc^2)^{-1}}$ and km~s$^{-1}$, respectively.
The best-fit relation is shown by the black dashed line in the top panel of Figure~\ref{fig:dust-aco}(a) and is consistent with the trend of the binned \aco~medians. The dispersion of data with respect to Equation~\ref{eqn_ew21_fit} is $\sigma \sim 0.12$~dex, and the standard deviation error returned by \texttt{curve\_fit} is $\pm$0.02 for both the fitted slope and intercept. We remind readers that the \aco~data here are converted from $\alpha_\mathrm{CO(2-1)}$ assuming $R_{21} = 0.65$, and thus it should be scaled by $R_{21}/0.65$ if $R_\mathrm{21}$ is known. If a SFR-dependent $R_{21}$ is used following \citetalias{2023arXiv231100407C}, the trend of \aco~in Figure~\ref{fig:dust-aco}(a) could be shallower by 30--40\%, as indicated by the red dashed line.

While the functional fit in Equation~\ref{eqn_ew21_fit} is based on $\Delta v$ measured at 150-pc scale, we also find a similar best-fit relation (dotted line) for six of those galaxies where $\langle \Delta v \rangle_\mathrm{90pc}$ is available. Because $\Delta v$ does not vary strongly between 90 and 150~pc scales \citep[see also][]{2022AJ....164...43S}, we would not expect this to change our results, and thus Equation~\ref{eqn_ew21_fit} should be applicable with $\Delta v$ measurements around 100-pc resolutions. We note that the evaluation of $\Delta v$ can also be affected by the number of gas components overlapping along the same sightlines, which could increase $\Delta v$ in barred galaxy centers. However, such effect is found to be mild (see \citetalias{2023ApJ...950..119T}, Appendix A), and we expect it to be even milder in our case, as $\langle \Delta v \rangle_\mathrm{150pc}$ is averaged over kpc-sized regions.

\subsection{Comparison to Previous Literature} \label{subsec:implication}

We compare our $\Delta v$-based prescription with existing \aco~prescriptions in the literature, including those based on metallicity \citep{2017MNRAS.464.3315A,2020ApJ...892..148S} or combining metallicity and stellar mass surface density (\citetalias{co-to-h2}). First, we investigate if metallicity alone could trace the observed \aco~variations. Figure~\ref{fig:dust-aco}(b) relates the measured \aco~with metallicity, using the same metallicity as those used in \citetalias{2023arXiv231100407C} to calculate \aco\ (see Section~\ref{subsec:data_aco}). 
The data and the power-law fit (black dashed line) overall agrees with the purely metallicity-dependent \aco~prescription from \citet{2017MNRAS.464.3315A} (orange dashed line)\footnote{The $Z'$ in the original prescription [$\alpha_\mathrm{CO} = 4.35 (Z')^{-1.6}$] was based on the PP04 calibration. Here we convert their prescription to the same (PG16S-based) metallicity scale as we adopt, using an approximate conversion based on \citet{2019A&A...623A...5D}.}, although the data scatter is larger than the trend with velocity dispersion. In the regime where our dataset is complete, the scatter of the observed \aco~is $\sigma \sim$0.1~dex with $\langle \Delta v \rangle_\mathrm{150pc}$ and 0.3~dex with $Z'$. This shows a significant improvement in predicting \aco~with our $\Delta v$-based prescription, compared to current metallicity-dependent prescriptions.
 
In the bottom panels of Figures~\ref{fig:dust-aco}(a) and (b), we relate the residuals of each \aco~fit with $Z'$ or $\langle \Delta v \rangle_\mathrm{150pc}$, in order to check if metallicity effects can explain any residual variation of \aco\ around the $\Delta v$ trend, or the opposite. Above the completeness limit, we find no trend between the residuals from the $\Delta v$ prescription and metallicity. On the other hand, the residuals from the metallicity fit clearly decrease with $\Delta v$ above the completeness threshold. This suggests that $\Delta v$ is crucial for tracing the \aco~changes, even without including metallicity effects. We have checked that the \aco~correlation with $Z'$ seen in this regime may come from the correlation between $Z'$ and $\Delta v$, as both variables decrease with the galactocentric radius.

Taking both metallicity and emissivity effects into account, \citetalias{co-to-h2} also suggested a tentative prescription\footnote{The original prescription included a molecular cloud surface density term which was assumed at 100~M$_\odot$~pc$^{-2}$. Here we adopt the same value and note that this helps avoid unrealistic \aco~values in low surface density regions (\citealt{2023ApJ...945L..19S}, \citetalias{2023ApJ...950..119T}).} based on \aco~measurements in nearby disks and (U)LIRGs:
\begin{equation}
\alpha_\mathrm{CO} \approx 2.9\ \exp\left(\frac{0.4}{Z'}\right) \left(\frac{\Sigma_\mathrm{star} + \Sigma_\mathrm{mol}}{100\ \mathrm{M_\odot\ pc^{-2}}}\right)^{-\gamma}~,
\label{eqn_alpha_B13}
\end{equation}
where $\gamma = 0.5$ if $\Sigma_\mathrm{star} + \Sigma_\mathrm{mol} > 100\ \mathrm{M_\odot}$~pc$^{-2}$ or $\gamma = 0$ otherwise. To compare the derived \aco~from our proposed prescription (Equation~\ref{eqn_ew21_fit}) with that from \citetalias{co-to-h2}, we apply both prescriptions to galaxies in the PHANGS sample (see Table~\ref{tab:sample}) using kpc-scale $Z'$ and $\Sigma_\mathrm{star}$. As we find $\Sigma_\mathrm{mol} \ll \Sigma_\mathrm{star}$ even with a (likely-overestimated) Galactic \aco~\citep{2022AJ....164...43S}, we neglect $\Sigma_\mathrm{mol}$ in Equation~\ref{eqn_alpha_B13}. We note that \citetalias{2023arXiv231100407C} also reported a similar \aco~relation that scales with $\Sigma_\mathrm{star}^{-0.5}$.

Figure~\ref{fig:compare-B13}(a) compares the \aco~values predicted by Equations~\ref{eqn_ew21_fit} and~\ref{eqn_alpha_B13}. Excluding the regime of $\log(\alpha_\mathrm{CO}) \gtrsim 0.65$ where \citetalias{co-to-h2} enforces a MW-like \aco~value with $\gamma = 0$ (which also corresponds to the low-confidence regime of our $\Delta v$-based prescription), the two prescriptions show an overall match with a $\sim$0.5 dex scatter. Despite a significant scatter, this general agreement may indicate that $\Delta v$ and $\Sigma_\mathrm{star}$ are tracing the same physical process that drives \aco\ variations. A likely scenario is that $\Delta v$ is set by the additional gravitational potential from stellar components, which can thus be tracked by $\Sigma_\mathrm{star}$ (see \citetalias{co-to-h2} and \citetalias{2023arXiv231100407C}). It is also possible that $\Delta v$ is a proxy of molecular gas surface densities and/or local CO intensities which could also reflect opacity and \aco~changes, as previous studies have found good correlations between these properties \citep[][see also Section~\ref{sec:discussion} for further discussion]{2022AJ....164...43S}. 

The \citetalias{co-to-h2} prescription was mostly based on \aco~measurements that were independent from ours and included several U/LIRGs in their sample, and the $\sim$0.5~dex scatter with our prescription is also consistent with the uncertainty estimated by \citetalias{co-to-h2}.
Therefore, the rough agreement seen in Figure~\ref{fig:compare-B13}(a) may also provide additional evidence for the validity of our proposed prescription. Compared to a $\Sigma_\mathrm{star}$-based prescription, one advantage of using a $\Delta v$-based prescription is that $\Delta v$ straightforwardly traces the optical depth change \citep{2022ApJ...925...72T,2023ApJ...950..119T}, making it closer to the underlying physics that could control \aco~variations. Another advantage is that $\Delta v$ can be directly obtained from the CO data. Thus, no ancillary multi-band data are needed to estimate \aco, which circumvents uncertainties in translating observations into $\Sigma_\mathrm{star}$. We remind readers that our prescription is calibrated to $\langle \Delta v \rangle_\mathrm{150pc}$ in CO\,(2--1), which is typically consistent with $\langle \Delta v \rangle_\mathrm{150pc}$ in CO\,(1--0) but may be different from that measured in higher-$J$ CO lines \citep{2020MNRAS.498.2440Y,2022ApJ...925...72T,2023ApJ...950..119T}. 
We also point out that systematic measurements of $\Delta v$ at 150-pc resolutions can be difficult across more extreme starbursts like U/LIRGs \citep[e.g.,][]{2019ApJ...882....5W}, which are usually more distant and/or more morphologically-disturbed than the galaxies in our sample. 

We note that the scaling of \aco~with $\Delta v$ in Equation~\ref{eqn_ew21_fit} is similar to what would be predicted by simple theoretical arguments. As shown by Equation~8 in \citet[see also related derivations in Chapter~19 of \citealt{2011piim.book.....D} and Chapter~8 of \citealt{2015arXiv151103457K}]{2020ApJ...903..142G}, the excitation temperature ($T_\mathrm{ex}$) under Large Velocity Gradient approximation with assumptions of a two-level optically-thick system can be written as
\begin{equation}
T_\mathrm{ex} \propto \rho_\mathrm{mol}\ \sqrt{\frac{L_\mathrm{mol} \cdot x_\mathrm{CO}}{\Delta v}}~,
\label{eqn_gong20}
\end{equation}
where $\rho_\mathrm{mol}$ and $L_\mathrm{mol}$ are the density and size of a CO-emitting molecular cloud, respectively. To first order, we also have $I_\mathrm{CO} \sim T_\mathrm{ex} \cdot \Delta v$ from the cloud. Thus, combining Equation~\ref{def_alphaCO} with Equation~\ref{eqn_gong20}, 
we obtain
\begin{equation}
\alpha_\mathrm{CO} = \frac{\Sigma_\mathrm{mol}}{I_\mathrm{CO}} \sim \frac{\rho_\mathrm{mol} \cdot L_\mathrm{mol}}{T_\mathrm{ex} \cdot \Delta v}
\propto \sqrt{\frac{L_\mathrm{mol}}{x_\mathrm{CO} \cdot \Delta v}}~.
\label{eqn_dv_theory}
\end{equation}
The resulting \aco~dependence on the inverse square root of $\Delta v$ is similar to the fits in Figure~\ref{fig:dust-aco}. While the fitted slope for $\langle \Delta v \rangle_\mathrm{150pc}$ (Equation~\ref{eqn_ew21_fit}) is slightly steeper than -0.5, we emphasize that the above calculation is highly simplified and is only for providing an intuitive check with theoretical expectations.

\subsection{Star Formation Efficiency in Galaxy Centers} \label{subsec:sfe_centers}

As \aco~determines the total molecular gas surface density ($\Sigma_\mathrm{mol}$, in units of M$_\odot$~pc$^{-2}$), the variation of \aco~directly affects the estimation of molecular gas depletion time ($t_\mathrm{dep}$) or SFE ($=1/t_\mathrm{dep}$):
\begin{equation} \label{eqn_sfe}
t_\mathrm{dep} = \Sigma_\mathrm{mol}\ /\ \Sigma_\mathrm{SFR} = \alpha_\mathrm{CO} \cdot I_\mathrm{CO}\ /\ \Sigma_\mathrm{SFR}~. \\
\end{equation}
While we examine only the SFE in this work, we note that the impact of \aco~on estimating the SFE per molecular cloud free-fall time is even more significant, as \aco~also affects the assessment of cloud density which changes the free-fall time \citep[e.g.,][]{2023arXiv231006501Q,2023ApJ...945L..19S}.
Motivated by the clear trend of galaxy centers having lower \aco~values (Figures~\ref{fig:dust-aco}(a) and~\ref{fig:compare-B13}(a)), we derive $t_\mathrm{dep}$ for the 12 galaxy centers with \aco~measurements (\citetalias{2023arXiv231100407C}; \citetalias{2023ApJ...950..119T}), using kpc-scale $\Sigma_\mathrm{SFR}$ and $I_\mathrm{CO(2-1)}$ (see Table~\ref{tab:sample}). Then, we examine how $t_\mathrm{dep}$ in galaxy centers derived from the measured \aco~would differ from that using the standard MW \aco~of 4.35 (or 6.7 in terms of $\alpha_\mathrm{CO(2-1)}$) $\mathrm{M_\odot\ (K~km~s^{-1}~pc^2)^{-1}}$. 

Figure~\ref{fig:compare-B13}(b) presents histograms of $t_\mathrm{dep}$ for the 12 galaxy centers. For the histogram using the measured \aco, we separate barred and non-barred galaxies with different colors. We find that the median $t_\mathrm{dep}$ with the MW \aco~is 4--5 times longer than that with the measured \aco. Furthermore, adopting the MW \aco~results in a similar $t_\mathrm{dep}$ of $\sim$3~Gyr between barred and non-barred centers. In contrast, if the measured \aco~is used, the median $t_\mathrm{dep}$ of barred and non-barred centers becomes 0.6~Gyr and 2.0~Gyr, respectively, differing by more than a factor of three. This suggests that SFE in barred galaxy centers tend to be higher than non-barred galaxy centers, and that using a constant \aco~can obscure such a trend.

\begin{figure*}
\begin{minipage}{.495\linewidth}
\centering
\includegraphics[width=\linewidth]{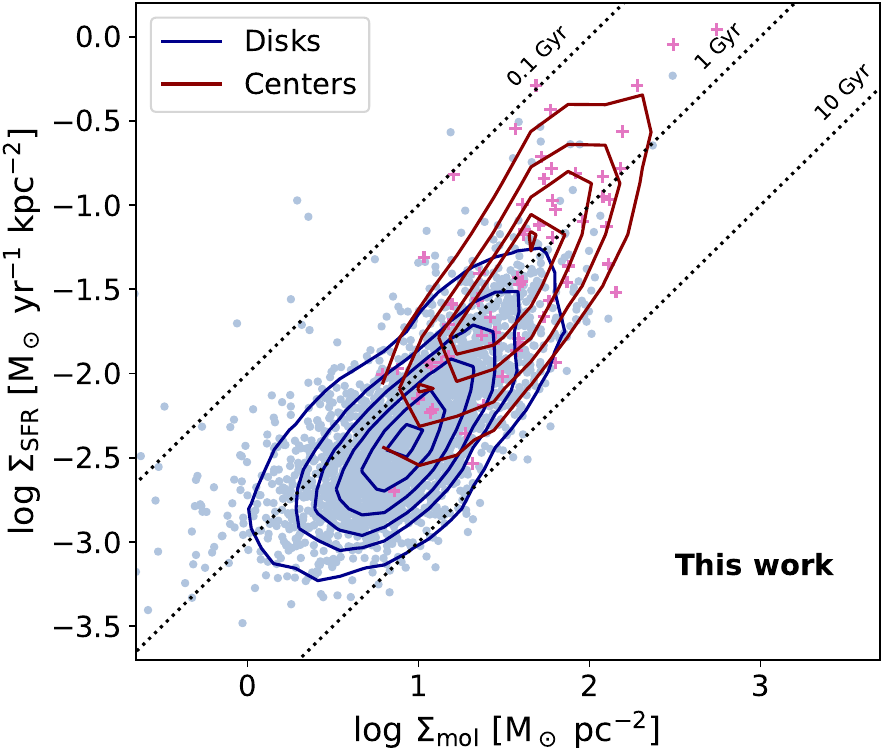}\\
(a)
\end{minipage}
\hfill
\begin{minipage}{.47\linewidth}
\centering
\includegraphics[width=\linewidth]{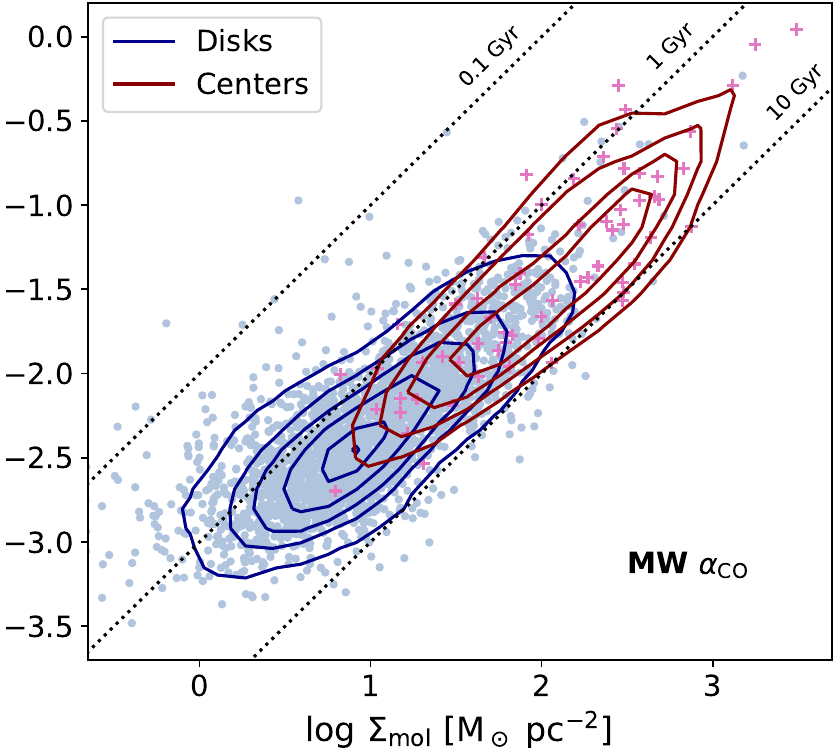}\\
(b)
\end{minipage}
\caption{The molecular Kennicutt-Schmidt (mKS) relation across 65 PHANGS galaxies, where the \aco~used to derive $\Sigma_\mathrm{mol}$ is based on (a) Equation~\ref{eqn_ew21_fit} or (b) the MW value. The thin dotted lines represent constant molecular gas depletion times ($t_\mathrm{dep}$) of 0.1, 1, and 10 Gyr. With our \aco~prescription, the galaxy centers clearly show a steeper trend than the disks, indicating shorter $t_\mathrm{dep}$ towards higher $\Sigma_\mathrm{mol}$. Adopting the MW \aco~instead results in a roughly constant $t_\mathrm{dep}$ for both centers and disks.}
\label{fig:ks-relation}
\end{figure*}

\begin{figure*}
\begin{minipage}{.32\linewidth}
\centering
\includegraphics[width=\linewidth]{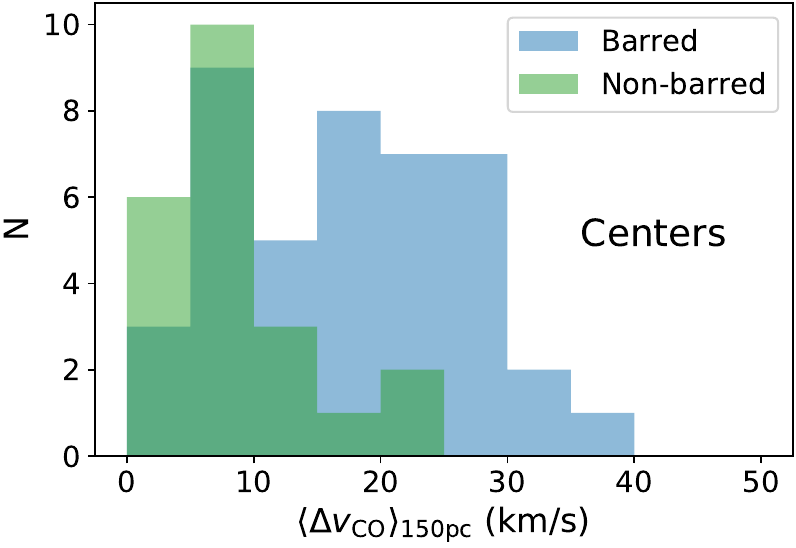}\\
\end{minipage}
\begin{minipage}{.34\linewidth}
\centering
\includegraphics[width=\linewidth]{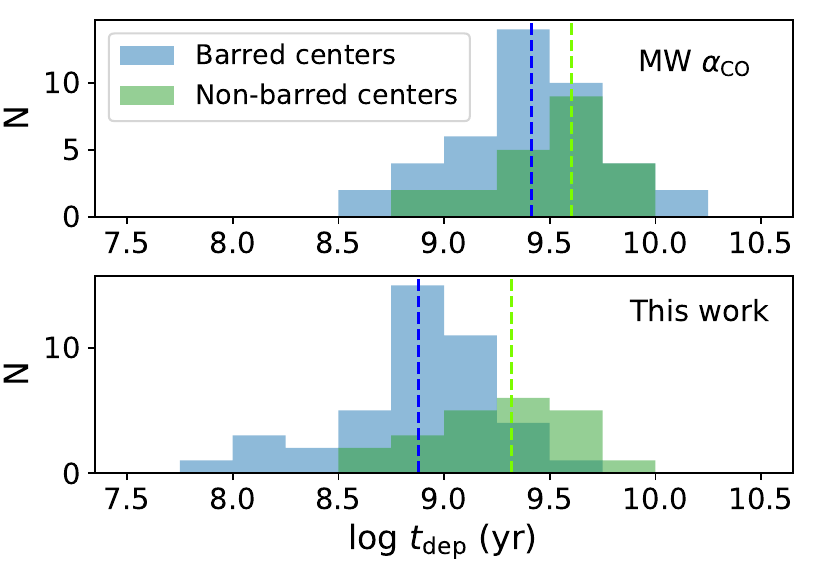}\\
\end{minipage}
\begin{minipage}{.34\linewidth}
\centering
\includegraphics[width=\linewidth]{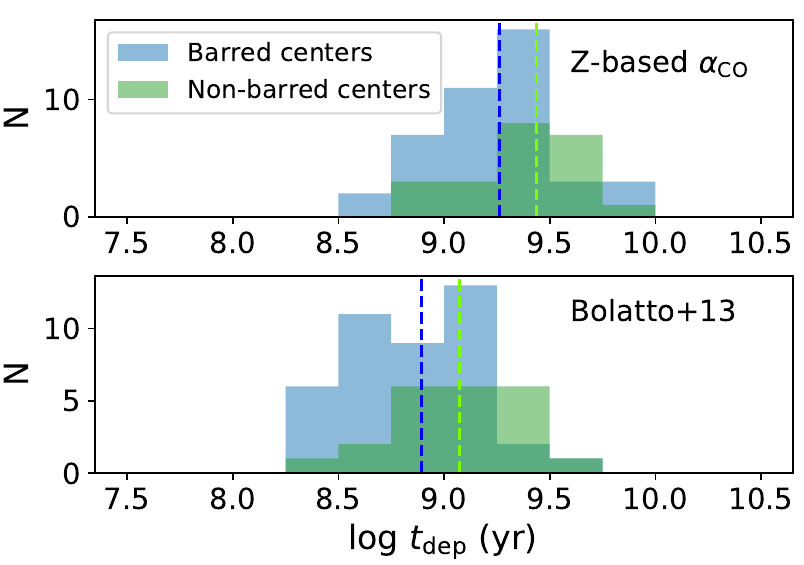}\\
\end{minipage} \\
\begin{minipage}{.33\linewidth}
\centering
\includegraphics[width=\linewidth]{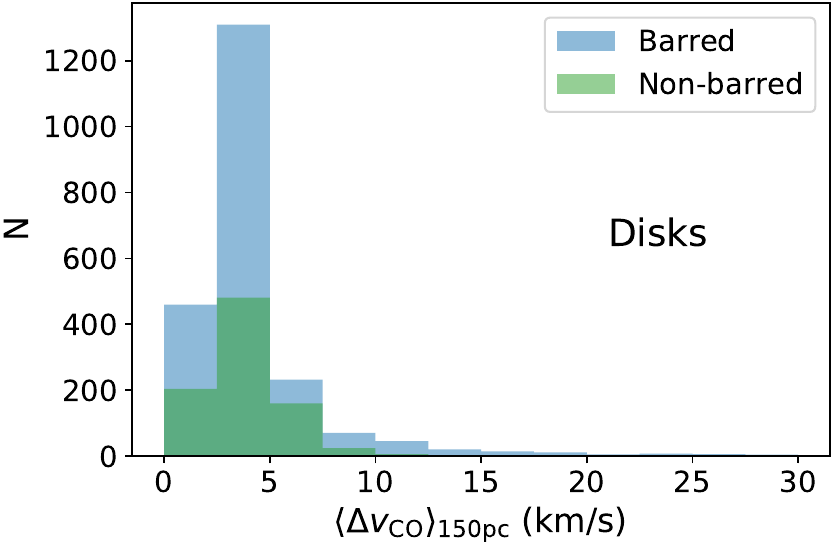}\\
\end{minipage}
\begin{minipage}{.33\linewidth}
\centering
\includegraphics[width=\linewidth]{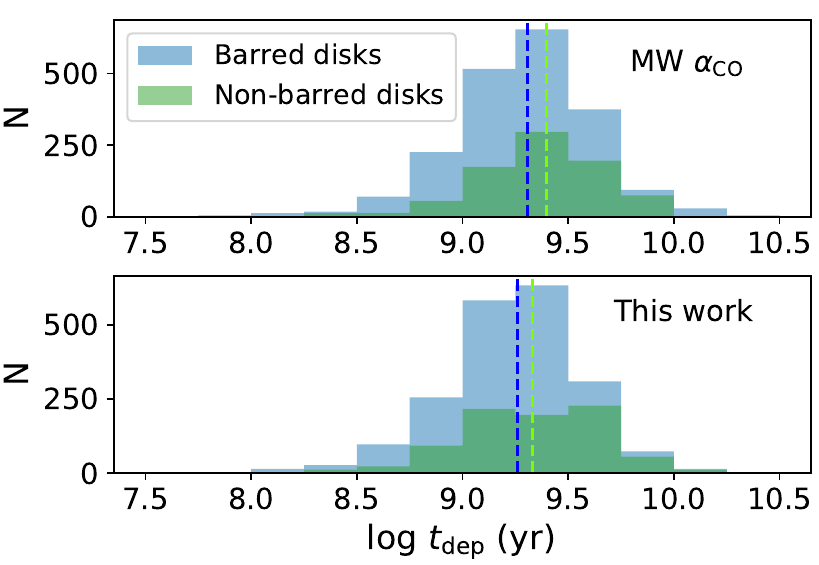}\\
\end{minipage}
\begin{minipage}{.34\linewidth}
\centering
\includegraphics[width=\linewidth]{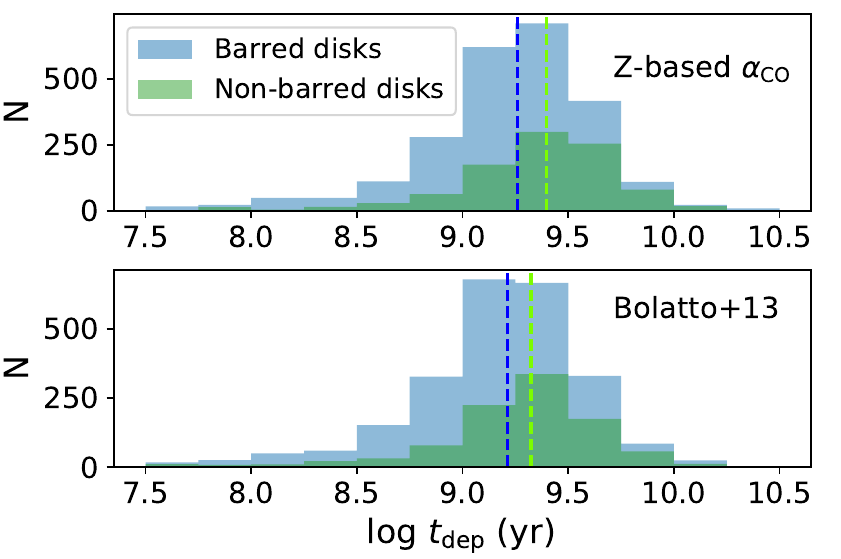}\\
\end{minipage}
\caption{Molecular gas velocity dispersion and the derived depletion time of PHANGS galaxies using four different \aco~prescriptions. The upper/lower panels show the centers/disks regions. The medians of the barred/non-barred distributions are indicated by the blue/green dashed lines. Our prescription reveals that barred centers tend to have higher star formation efficiency than non-barred centers due to a generally higher velocity dispersion, but such trend is easily obscured using other prescriptions. Contrary to the centers, the disk regions show consistent distribution of velocity dispersion and depletion time between barred and non-barred galaxies, regardless of which prescription is used.}
\label{fig:all-phangs}
\end{figure*}

\subsection{Systematic Impact on Star Formation Efficiency} \label{subsec:sfe}

With the PHANGS sample (Table~\ref{tab:sample}), we investigate the impact of different \aco~prescriptions on SFE or $t_\mathrm{dep}$ in the centers and disks of barred and non-barred galaxies. Figure~\ref{fig:ks-relation} shows the molecular Kennicutt-Schmidt (mKS) relation across all 65 galaxies measured at the 1.5-kpc scale, comparing $\Sigma_\mathrm{mol}$ determined from our \aco~prescription (Equation~\ref{eqn_ew21_fit}) with that determined using a MW \aco. It is clear that adopting the MW \aco~results in a wider range of $\Sigma_\mathrm{mol}$ with values reaching $> 1000$~M$_\odot$~pc$^{-2}$ in galaxy centers, while our prescription suggests $\Sigma_\mathrm{mol} < 200$~M$_\odot$~pc$^{-2}$ in general. Furthermore, our prescription reveals a trend of higher SFE towards higher $\Sigma_\mathrm{mol}$, which steepens the mKS relation for galaxy centers and other high-$\Sigma_\mathrm{mol}$ regions. With the MW \aco, however, both galaxy centers and disks exhibit a roughly constant SFE. These results show that \aco~and $\Sigma_\mathrm{mol}$ in galaxy centers may overall be overestimated by a factor of 5 with the MW \aco, and that the choice of \aco~greatly affects our understanding of galactic-scale star formation.  

Figure~\ref{fig:all-phangs} presents histograms of velocity dispersion and $t_\mathrm{dep}$ across the PHANGS sample, separating centers (upper panels) and disks (lower panels) for barred (blue) and non-barred (green) galaxies. In non-barred galaxy centers, $\langle \Delta v \rangle_\mathrm{150pc}$ is typically $< 10$ km~s$^{-1}$, while barred centers span a significantly wider range up to $\sim$40 km~s$^{-1}$. On the other hand, barred and non-barred disks show consistent velocity dispersion, with $\langle \Delta v \rangle_\mathrm{150pc}$ typically below 5~km~s$^{-1}$ but reaching up to 10~km~s$^{-1}$. These distributions agree with \citet{2020ApJ...901L...8S}, who reported similar $\Delta v$ between galaxy disks and non-barred centers but an overall $\sim$5 times higher $\Delta v$ in barred centers. 

We then examine the distribution of $t_\mathrm{dep}$ derived with different \aco~prescriptions. Using our $\Delta v$-based prescription, we find distinctly different $t_\mathrm{dep}$ between barred and non-barred centers, with the mean/median of $t_\mathrm{dep}$ in barred galaxy centers ($\sim$700~Myr) being 3 times shorter than in non-barred centers (${\sim}2.1$~Gyr). The 16th--84th percentile ranges for $t_\mathrm{dep}$ in barred and non-barred centers is 0.3--1.6 and 0.8--3.6~Gyr, respectively. In contrast, all other prescriptions result in $< 0.2$~dex difference between the median $t_\mathrm{dep}$ of the two types of systems. Such a small difference between barred and non-barred centers is even true for the \citetalias{co-to-h2} prescription which shows similarly short $t_\mathrm{dep}$ for all galaxy centers that generally matches our results. Particularly, the MW \aco~leads to two completely overlapping $t_\mathrm{dep}$ distributions, overestimating the overall $t_\mathrm{dep}$ in barred galaxy centers by a factor of 3--4 if compared to our results. As for the disks, the median of $t_\mathrm{dep}$ remains consistent at 2--3~Gyr across all four prescriptions, while it is found to be systematically lower in barred galaxies than in non-barred galaxies by $\sim$0.1~dex.

Notably, our prescription reveals short $t_\mathrm{dep}$ down to $\lesssim$100~Myr in some barred galaxy centers, which is not seen with other prescriptions. Such a short time scale is supported by recent simulations of galaxy centers including effects from bar-driven inflows \citep[e.g.,][]{2019MNRAS.490.4401A,2020MNRAS.497.5024S,2021ApJ...914....9M}. In addition, 
we note that the overall $t_\mathrm{dep}$ for galaxy centers is similar between our result and \citetalias{co-to-h2}'s, both suggesting $t_\mathrm{dep} \sim$1~Gyr which is shorter than the disks value of $\sim$3~Gyr. This factor-of-three difference between centers and disks is consistent with recent simulations \citep[e.g.,][]{2020MNRAS.492.2973T}. However, using the MW or $Z$-based \aco~for galaxy centers obscures such difference and leads to similar $t_\mathrm{dep}$ across entire galaxies.

\section{Discussion} \label{sec:discussion}

The correlation of \aco~with $\sim$100-pc scale velocity dispersion with only a $\sigma \sim 0.1$~dex scatter (see Section~\ref{subsec:prescription}), contrary to $\sim$0.3~dex or larger scatter using $Z$- and/or $\Sigma_\mathrm{star}$-based prescriptions, shows that velocity dispersion is an excellent observational tracer for \aco~variations in star-forming galaxies. The rationale behind such a strong relation may be that $\Delta v$ directly traces the optical depth changes that are the dominant effect responsible for altering \aco~across these galaxies, as it has been shown that opacity variation is the primary driver of \aco~in various galaxy centers \citep{2020A&A...635A.131I,2022ApJ...925...72T,2023ApJ...950..119T}. However, effects of CO-dark gas and CO excitation can also be important to explain \aco~variations across the galaxy disks, which have therefore motivated previous \aco~prescriptions based on metallicity and/or CO integrated intensity \citep[e.g.,][]{2012MNRAS.421.3127N,2015A&A...583A.114H,2016A&A...588A..23A,2017MNRAS.464.3315A,2020ApJ...903..142G}. 

As discussed in Section~\ref{subsec:implication}, the correlation of \aco~with metallicity ($Z'$) is indirectly included in the dependence with $\Delta v$ because both $Z'$ and $\Delta v$ vary with galactocentric radius and are thus correlated. Furthermore, statistical studies on molecular cloud properties have shown that velocity dispersion also correlates well with molecular gas surface density and the CO integrated intensity across galaxy disks \citep{2009ApJ...699.1092H,2020ApJ...901L...8S,2022AJ....164...43S,2021MNRAS.502.1218R}. Therefore, it is likely that our $\Delta v$-based prescription contains opacity variations and metallicity gradients as well as the physics of the \aco--$I_\mathrm{CO}$ correlation suggested by simulation studies \citep{2012MNRAS.421.3127N,2020ApJ...903..142G,2022ApJ...931...28H}. This means that the proposed prescription (Equation~\ref{eqn_ew21_fit}) may incorporate more than one piece of physics into a single scaling relation, which could explain why the trend holds across different galactic environments.
We also note that metallicity effects on \aco~should be more drastic in low-metallicity dwarf galaxies due to the lack of dust shielding that can prevent CO from dissociation, and thus metallicity variations being included in our $\Delta v$-based prescription might only be true in the context of MW-like star-forming disk galaxies as represented by the PHANGS sample.

In Sections~\ref{subsec:sfe_centers} and \ref{subsec:sfe}, our prescription (based on the dust \aco~measurements) suggests lower \aco~in barred galaxy centers that lead to higher SFE than non-barred centers and the disks. This low \aco~and high SFE in barred centers imply that the amount of molecular gas can be overestimated by previous studies due to inaccurate \aco~or the assumption of a constant SFE. By comparing the derived $\Sigma_\mathrm{mol}$ under different \aco~assumptions for all galaxies in Table~\ref{tab:sample}, we find that the median $\Sigma_\mathrm{mol}$ of barred centers is 3 times higher than that of non-barred centers if using a MW-like \aco. On the other hand, our \aco~prescription results in only 1.3 times higher $\Sigma_\mathrm{mol}$ in barred centers.
Therefore, it is likely that the enhanced SFE is a more important factor causing high SFR observed in barred galaxy centers, compared to an increased amount of molecular gas driven inwards by bars. 

Recent studies using \aco~prescriptions from \citet{2012MNRAS.421.3127N} or \citetalias{co-to-h2} also show that barred galaxies tend to have higher central gas concentration than non-barred galaxies, although the degree of concentration is not as significant as using a constant \aco\ \citep{1999ApJ...525..691S,2005ApJ...632..217S,2006ApJ...649..181S,2007PASJ...59..117K}. 
Such accumulation of gas towards the centers can increase SFR in barred centers, and it is consistent with the theoretical expectation that non-axisymmetric gravitational potential from bars can induce gas inflows and transport more gas into galaxy centers \citep[e.g.,][]{1995MNRAS.277..433W,2004ApJ...600..595R,2012ApJ...747...60K,2020MNRAS.499.4455T}. Bars thus influence the secular evolution of galaxies by redistributing molecular gas mass and angular momentum \citep[see review by][]{2004ARA&A..42..603K}. 

Studies have also shown that if \aco~changes were treated properly, starbursts in galaxy centers and variations of SFRs across nearby galaxies are primarily driven by higher SFE rather than increased molecular gas fraction \citep{2013AJ....146...19L,2020MNRAS.492.6027E,2020MNRAS.493L..39E,2023arXiv230203044D}. This is contrary to studies using constant, $Z$-based, or $\Sigma_\mathrm{star}$-based \aco, which resulted in similar SFE between barred and non-barred galaxies \citep[e.g.,][see also Section~\ref{subsec:sfe}]{2012ApJ...758...73S,2021A&A...656A.133Q}. With our proposed \aco~prescription, we find enhanced SFE in barred centers, which could originate from variations in molecular gas distribution, density structure, or dynamical effects of turbulence and shocks powered by stellar feedback \citep[e.g.,][]{2009A&A...508L..35K,2012ApJ...760L..16R,2023ApJ...944L..19L}. However, these factors driving SFE variations are the same ones that can alter CO emissivity and \aco. 
Therefore, only with accurate \aco~values can we disentangle SFE from \aco~and unravel the physical drivers of SFR. Using the latest and best possible measurements of \aco~and molecular gas properties across a sample of nearby galaxies, our work lays a foundation for benchmarking \aco~calibration in star-forming galaxies (including starbursting galaxy centers) and allows for further investigation on SFE, SFE per cloud free-fall time, or other related properties that can improve our knowledge of galaxy evolution.

\section{Conclusions} \label{sec:conclusion}

We construct a new \aco~prescription applicable to star-forming galaxies, where CO emissivity variations are critical in altering \aco. The prescription is a major step towards precise calibration of \aco~across galaxies, and it reveals unprecedented trends in star formation properties which may have been obscured by previous \aco~prescriptions. Our key results are summarized as follows:

\begin{enumerate}

\item The strong anti-correlation between measured \aco~and CO velocity dispersion ($\Delta v$) at $\sim$100-pc scales shows that $\Delta v$ is useful for predicting \aco, and it suggests that CO opacity altered by $\Delta v$ changes or other correlated properties of the molecular gas across the entire galaxies are primary drivers of \aco~in star-forming galaxies.   

\item The proposed \aco~prescription (Equation~\ref{eqn_ew21_fit}) is applicable to regions with metallicity above $0.6\ Z_\odot$ and $\langle \Delta v \rangle_\mathrm{150pc} \gtrsim 3$ km~s$^{-1}$. The expected scatter in \aco~is $\sigma \sim 0.1$~dex, which is a substantial improvement over existing \aco~prescriptions. Our $\Delta v$-based prescription has the advantage of connecting directly to the physical causes of \aco~change (e.g., CO opacity) as well as requiring only the CO observations which is most relevant to tracing molecular gas.  

\item With the measured \aco, we find distinctly shorter molecular gas depletion time ($t_\mathrm{dep}$) in barred galaxy centers than non-barred galaxy centers, as well as a generally shorter $t_\mathrm{dep}$ in galaxy centers than the disks. In contrast, assuming a constant MW \aco~results in $t_\mathrm{dep} \sim3$~Gyr for all regions, which underestimates the star formation efficiency (SFE) in galaxy centers and also obscures the difference between barred and non-barred galaxies.  

\item Our prescription reveals short $t_\mathrm{dep}$ down to 100 Myr in barred galaxy centers, with the median $t_\mathrm{dep}$ (0.7$^{+0.9}_{-0.4}$~Gyr) being 3 times shorter than in non-barred galaxy centers (2.1$^{+1.5}_{-1.3}$~Gyr). However, all other prescriptions (MW, metallicity-based, and \citetalias{co-to-h2}) show $< 0.2$~dex difference between the two regions, even if \citetalias{co-to-h2} results in an overall shorter $t_\mathrm{dep}$ for galaxy centers which aligns better with our results. Thus, SFE in barred galaxy centers may be underestimated by a factor of three or more in previous studies due to \aco~uncertainties. 

\item All four prescriptions tested in this work show similar $t_\mathrm{dep}$ of 2--3~Gyr in the disk regions and non-barred galaxy centers across the PHANGS sample, which is in good agreement with previous literature \citep[e.g.,][]{2008AJ....136.2782L,2011MNRAS.415...61S,2023ApJ...945L..19S}.

\end{enumerate}

{
This work was carried out as part of the PHANGS collaboration. 
Y.-H.T. and K.S. acknowledge funding support from NRAO Student Observing Support Grant SOSPADA-012 and from the National Science Foundation (NSF) under grant No. 2108081.
Y.-H.T. also acknowledges support by the Ministry of Education of Taiwan through Government Scholarship to Study Abroad.
J.S. acknowledges support by the Natural Sciences and Engineering Research Council of Canada (NSERC) through a Canadian Institute for Theoretical Astrophysics (CITA) National Fellowship. 
KG is supported by the Australian Research Council through the Discovery Early Career Researcher Award (DECRA) Fellowship (project number DE220100766) funded by the Australian Government. 
KG is supported by the Australian Research Council Centre of Excellence for All Sky Astrophysics in 3 Dimensions (ASTRO~3D), through project number CE170100013.
HAP acknowledges support by the National Science and Technology Council of Taiwan under grant 110-2112-M-032-020-MY3.
MB gratefully acknowledges support from the ANID BASAL project FB210003 and from the FONDECYT regular grant 1211000. 
IC thanks the National Science and Technology Council for support through grant 111-2112-M-001-038-MY3, and the Academia Sinica for Investigator Award AS-IA-109-M02.
JC acknowledges funding from the Belgian Science Policy Office (BELSPO) through the PRODEX project “JWST/MIRI Science exploitation” (C4000142239).
MC gratefully acknowledges funding from the DFG through an Emmy Noether Research Group (grant number CH2137/1-1).
COOL Research DAO is a Decentralized Autonomous Organization supporting research in astrophysics aimed at uncovering our cosmic origins.
SKS acknowledges financial support from the German Research Foundation (DFG) via Sino-German research grant SCHI 536/11-1.
AU acknowledges support from the Spanish grant PID2019-108765GB-I00, funded by MCIN/AEI/10.13039/501100011033. 
J.~D.~H. gratefully acknowledges financial support from the Royal Society (University Research Fellowship; URF/R1/221620).

This paper makes use of the following ALMA data: \linebreak
ADS/JAO.ALMA\#2012.1.00650.S, \linebreak 
ADS/JAO.ALMA\#2013.1.00803.S, \linebreak 
ADS/JAO.ALMA\#2013.1.01161.S, \linebreak 
ADS/JAO.ALMA\#2015.1.00925.S, \linebreak 
ADS/JAO.ALMA\#2015.1.00956.S, \linebreak 
ADS/JAO.ALMA\#2017.1.00392.S, \linebreak 
ADS/JAO.ALMA\#2017.1.00766.S, \linebreak 
ADS/JAO.ALMA\#2017.1.00886.L, \linebreak 
ADS/JAO.ALMA\#2018.1.01321.S, \linebreak 
ADS/JAO.ALMA\#2018.1.01651.S. \linebreak 
ADS/JAO.ALMA\#2018.A.00062.S. \linebreak 
ALMA is a partnership of ESO (representing its member states), NSF (USA), and NINS (Japan), together with NRC (Canada), NSC and ASIAA (Taiwan), and KASI (Republic of Korea), in cooperation with the Republic of Chile. The Joint ALMA Observatory is operated by ESO, AUI/NRAO, and NAOJ. The National Radio Astronomy Observatory is a facility of the National Science Foundation operated under cooperative agreement by Associated Universities, Inc.
This work is based in part on data products from the Wide-field Infrared Survey Explorer\citep{2010AJ....140.1868W}, which is a joint project of the University of California, Los Angeles, and the Jet Propulsion Laboratory/California Institute of Technology, funded by the National Aeronautics and Space Administration.
We acknowledge the usage of the SAO/NASA Astrophysics Data System.


\facilities{ALMA, WISE, Herschel, VLA}

\software{\texttt{matplotlib} \citep{Hunter:2007}, \texttt{numpy} \citep{harris2020array}, \texttt{scipy} \citep{Virtanen_2020}, \texttt{astropy} \citep{2013A&A...558A..33A,2018AJ....156..123A}, \texttt{ipython} \citep{PER-GRA:2007} }
}


\bibliography{reference}{}
\bibliographystyle{aasjournal}



\end{document}